# The influence of the curvature dependence of the surface tension on the geometry of electrically charged menisci

**Ramiro dell'Erba**[1], **Francesco dell'Isola**[2], **and Giacomo Rotoli**[3]

[1] IRTEMP CNR, Via Toiano 6, Arco Felice 80072 (NA)
[2] Dipartimento di Ingegneria Strutturale e Geotecnica, Università di Roma La Sapienza, Via Eudossiana 18, I-00184 Roma, Italy
[3] Dipartimento di Energetica, Università dell' Aquila, Roio Poggio, I-67040 L'Aquila, Italy

We evaluate how the curvature dependence of surface tension affects the shape of electrically charged interfaces between a perfectly conducting flui and its vapour. We consider two cases: i) spherical droplets in equilibrium with their vapour; ii) menisci pending in a capillary tube in presence of a conducting plate at given electric potential drop.
Tolman-like dependence of surface tension on curvature becomes important when the "nucleation radius" is comparable with the interface curvature radius. In case i) we prove existence of the equilibrium minimal radius and estimate its dependence on the electric field and Tolman-like curvature effects. In case ii) the menisci are subject to the gravitational force, surface tension and electrostatic fields We determine the unknown surface of the menisci to which the potential is assigned using an iterative numerical method and show that Tolman-like corrections imply:
1) a variation of the height (up to 10% in some cases) of the tip of the menisci;
2) a decrease of the maximum electrical potential applicable to the menisci before their breakdown amounting to $40V$ over $800V$ in the considered cases.
We conjecture that these effects could be used in new experiments based on electric measurements to determine the dependence of the equilibrium surface tension on curvature.

## 1 Introduction

In Tolman [1] the classical thermodynamic theory of interfaces between different phases due to Gibbs [2] (see Scientifi Papers (1948)) was applied to study the equilibrium of spherical drops with their interfaces. In particular Tolman proved, in the framework of Gibbs' theory of excess thermodynamic quantities, that the second principle of thermodynamics implies an exponential dependence at equilibrium of the surface tension on the curvature of interphase surface. Moreover, he evaluated the characteristic radius for this exponential decay towards a constant value in terms of known thermodynamic quantities. La Mer and Pound [3] demonstrated that Tolman's formulas fail in predicting experimental evidence. Several efforts were subsequently performed to generalize Gibbs' theory to match quantitative experimental data. For a detailed report of these efforts we refer to Fisher and Israelachvili [4]. We quote here the treatments applied by Defay and Prigogine [5], Alts and Hutter [6] who consistently used the concept of excess quantities, and the results obtained by dell'Isola [7] using the concept of two-dimensional non-material shell-like Defay-Prigogine interface.

However, to our knowledge, by only using a three-dimensional second gradient theory, such as that of Germain [8], to model the material behavior of the interface between different phases (see dell'Isola and Rotoli [9], dell'Isola et al. [10], dell'Isola et al. [11]), it is possible to adjust the parameters which control

---

*Roberto Stroffolini in Memoriam: To the Maestro who introduced us to Continuum Mechanics and Electrodynamics*

the dependence of the equilibrium surface tension on curvature and to fi the available experimental data (see Fisher and Israelachvili [4]).

The fact that the nucleation energy of microscopic bubbles is lower than those energies that are computed with a constant surface tension is generally accepted, but there is no clear experimental evidence that this phenomenon is related only to a constitutive dependence of surface tension on curvature. We explicitly remark that in a theory in which surface tension is given as a constitutive equation in terms of temperature and curvature the second principle of thermodynamics implies that surface tension is a decreasing function of the curvature radius. This last result is in contradiction with the aforementioned property of nucleation energy. As a consequence we expect that a more sophisticated thermodynamic theory for surface tension must be formulated (e.g. using the methods developed by dell'Isola and Kosinski [12]) in which a more refine kinematic description of the interface is introduced: this theory, applied to equilibrium states, will lead to a relation between equilibrium surface tension and curvature which we call Tolman-like correction.

In this paper, in order to obtain further insight into these problems and make possible a comparison among different dependencies of equilibrium surface tension on curvature, we consider their influenc on the shape of electrically charged interfaces. The only physical quantities which we will consider are electrical surface charge density and surface tension. We expect that the most important phenomena occurring at the interface between non-polar charged fluid will be described in terms of them.

We do not assume any constitutive influenc on the surface tension by the electric fiel itself. A simple estimate of the energy density of the interface and the electric fiel energy shows that this effect must be considered only for very strong electric fields typically larger than the dielectric breakdown of the fluid

We shall address two cases: i) a conducting droplet surrounded by its (non-conducting) vapour; ii) a conducting meniscus in a typical electrocapillarity experiment: a conducting truncated conical pipet, on which the meniscus is pendent, is placed opposite to a conducting plane while a potential difference is applied between them. The firs case is considered in order to supply a simple benchmark case to the analysis developed in ii). Indeed the spherical symmetry allows an easy determination of the electrostatic field involved. The investigation of the equilibrium of drops and bubbles for electrically charged conducting interfaces generalizes the classical results found in a rational form by Romano [13]. Case ii) was investigated by Joffre [14], by assuming the surface tension $\gamma$ to be constant. Our aim is to extend these results to menisci of which the curvature mean radius is locally comparable with the minimal nucleation radius as define in dell'Isola et al. [10] so that the equilibrium Tolman-like dependences of surface tension on curvature must be considered.

The main difficult is the determination of the electric fiel close to the interfaces. However, by using the methods outlined in Sect. 3 we are able to generalize the results obtained in Joffre [14] for the conical-shaped capillary tube and constant surface tension (i.e., the case of large radius of capillary tube). When the Tolman-like dependence of the surface tension, $\gamma$, on curvature is important (i.e. in the neighbourhood of the critical temperature and/or minimal nucleation radius) we obtain relevant differences to the case of constant $\gamma$ in predicting the break-down electrical potential: the choice of the Tolman-like dependence can lead to its decrease by about 5%. Moreover, the height of the tip of the menisci is increased by about 7% when surface tension is increasing with increasing curvature radii or decreased by about 30% when surface tension is decreasing with increasing radii.

We believe that such effects could be used to support a new experimental measure of the dependence of equilibrium surface tension on curvature simply by measuring the electric potentials in an apparatus with small-radii-capillary tubes. These experiments could help to choose, among the proposed Tolman-like corrections, those amendments which best match the measurements. This choice is a vexed question much discussed since Tolman firs published his paper in 1949 [1]. Indeed:

    -very few experiments could be performed to settle it (see Fisher and Israelachvili [4]),

    -much of the evidence is only indirect (La Mer and Pound [3], Kumar et al. [15]),

    -several inferences used in these indirect measurements are doubtful (dell'Isola and Rotoli [9]).

We conclude by underlining that the geometry of our problem slightly differs from that chosen by Joffre [14]. Indeed, in order to avoid singularities in the electric fiel in the neighbourhood of the sharp edges shared by the meniscus and the capillary tube we choose to use as capillary tubes only truncated cones.

## 2 Equilibrium of fluid phases separated by electrically charged spherical interfaces

In this section we generalize to the case of electrically charged spherical interfaces a theorem proved by Romano [13] starting from the following balance equations:

$$\begin{aligned} \nabla p &= 0 \quad \text{in} \quad C_1 \cup C_2, \\ \gamma &= \hat{\gamma}(\vartheta), \\ \frac{2\gamma}{R} &= [\![p]\!] \quad \text{on} \ S, \\ [\![g(p)]\!] &= 0 \quad \text{on} \ S. \end{aligned} \quad (1)$$

Where $p$ is the pressure; $\gamma$ the surface tension; $R$ the drop radius (coincident with the radius of curvature); $[\![\cdot]\!]$ denotes the jump of the bracketed quantity across $S$ (i.e. its limit value from inside $S$ minus that from outside); $g(p)$ is the density of Gibbs energy (chemical potential) and $C_1$ and $C_2$ the inner and the outer regions of the bubble while $S$ is the interface surface. In (1) the volume body forces are neglected while the temperature fiel $\vartheta$ is assumed to be constant in space. The aforementioned theorem states the following:

**Theorem 1** *If the temperature is given in the interval $(\theta_*, \theta_c)$, then the vapour at uniform pressure $p_v \in (p_0, p_2)$ is at equilibrium with drops of the radius $R$ of its liquid if and only if the liquid has a suitable pressure $p_l \in (p_0, p_l^*)$ and $R \in [2\gamma/(p_l^* - p_2), \infty]$. As a consequence there exists an equilibrium minimal radius given by $R_{min} = 2\gamma/(p_l^* - p_2)$*

We used the following notation:

$\theta_*$ = triple point temperature,
$\theta_c$ = critical temperature,
$p_1$ = spinodal pressure,
$p_v$ = vapour pressure,
$p_0$ = unique solution of $g_v(p_0) = g_l(p_1)$,
$g_v$ and $g_l$ are the Gibbs energies of the vapour and the liquid respectively,
$p_2$ = saturation pressure,
$p_l^* = g_l^{-1}(g_v(p_2))$.

Consider now a conducting drop carrying an electric charge $Q$. We assume it is subject to electrostatic conditions and introduce the radial component $E \propto Q/r^2$ ($r$ being the distance from the center of the drop) of the electrostatic field Introducing into the balance equations a Tolman-like correction for surface tension $\gamma(R_g) = \gamma_0 f(R_g)$, $R_g$ being the surface mean curvature radius (coincident in this section with the drop radius), and $f(R_g)$ a regular function to be specified (1) become

$$\begin{aligned} \nabla p &= 0 \quad \text{in} \quad C_1 \cup C_2, \\ \gamma &= \gamma_0 f(R_g), \\ \frac{2\gamma(R_g)}{R_g} + E\sigma &= [\![p]\!], \\ [\![g(p)]\!] &= 0 \quad \text{on} \ S. \end{aligned} \quad (2)$$

in which $\sigma$ is the surface charge density at the liquid-vapour interface.

It should be noted that firs the condition $[\![g(p)]\!] = 0$ is obtained when the surface Gibbs energy $g_\sigma$ in the local form of jump conditions expressing balance of energy is neglected (for more details see Alts and Hutter [6] or dell'Isola and Romano [16]); second we assume that the drop does not discharge itself by air ionization; third in the present section for the functional form of the Tolman-like correction the relations

$$\gamma_1(R_g) = \gamma_0 \left(1 - e^{-\frac{R_g}{R_0}}\right), \quad (3)$$

$$\gamma_2(R_g) = \gamma_0 \frac{1}{2}\left[1 + \tanh\left(\frac{R_g - R_0}{\alpha R_0}\right)\right] \quad (4)$$

are chosen. In these the so-called "minimal nucleation radius" $R_0$ appears (cf. dell'Isola et al. [10]). $R_0$ is a temperature dependent quantity, strictly related to microscopic parameters of the considered fluid as e.g. the molecular dimensions. The dependence of $R_0$ on temperature is of the form of a standard critical power law, i.e., $R_0 = \mathscr{L}(1 - T/T_c)^\beta$ with $\beta < 0$ where $\mathscr{L}$ is a length of the order of magnitude of molecular dimension. The firs expression is similar to Tolman's original formula except for the definitio of and therefore the values to be assigned to $R_0$; the second one is an empirical formula. It approximates experimental data better and is based on the results obtained by dell'Isola et al. [10]: the constitutive constant $\alpha$ is related to the state equation describing the microscopic behaviour of the considered material.

We emphasize that the introduced Tolman-like corrections cannot be regarded as constitutive equations for surface tension. Indeed it is well-known that the second principle of thermodynamics does not allow surface tension to depend on curvature unless the interface between phases is modelled by a micro-structured 2D-continuum. In order to develop a thermodynamically coherent model one should introduce in the constituive equations at least the surface gradient of curvature and extra kinematical descriptors and must then model the variation of the thickness of the interface and the surface mass density together with their corresponding evolution equations. An alternative possibility to force thermodynamic compatibility between Tolman-like corrections and free energy is to assume an explicit dependence of the latter on curvature and to introduce, as in the theory of shells, couple stresses.

Therefore, we regard Tolman-like corrections to be valid only for equilibrium conditions, and we postpone to more general treatments the formulation of constitutive equations for surface tension locally valid under non-equilibrium conditions. Such constitutive equations should be compatible with the request of a decreasing dependence of surface free energy on the radius of curvature: indeed a non-decreasing dependence would lead to ill-posed minimization problems and the related relaxed problems would no longer involve surface tension.

These corrections can be used whenever the radius of the drop is larger than $R_0$. We note that a drop of radius $R \sim 2R_0$ can already be considered sufficientl far from nucleation as described in dell'Isola et al. [11]. Otherwise close to $R_0$ the non-local nature of the process must be considered, and a corrected form of the free energy must be introduced e.g. suggested in dell'Isola et al. [11].

Equation (2) implies that in every phase the pressure fiel is constant. Therefore (once the temperature, the charge $Q$ or alternatively the potential $U$ of the drop and the vapour pressure $p_v$ are fixed the set of (3), (4) becomes an algebraic system in the unknowns $R$ and $p_l$.

In order to formulate the generalization of theorem 1 to the case represented by (2), (5) we introduce the following potential function

$$LC_U(R) := \frac{2\gamma(R)}{R} + \frac{U^2}{\varepsilon R^2}, \qquad (5)$$

where $\varepsilon$ is the dielectric constant of the vapour and the second term is obtained by recalling the expression for the electrostatic fiel generated by an isolated uniform surface charge. Simple calculations prove that for both $\gamma_i$ introduced in (6), (7) the functions $\gamma_i(R)/R$ are decreasing with $R$. More generally, this property has been shown for a large class of continuum models of flui interfaces from the second principle of thermodynamics (see dell'Isola et al [11] and dell'Isola [7]). Therefore the *equilibrium minimal radius* $R_{min}$ can be define by

$$R_{min} := (LC_U)^{-1}(\max(p_l - p_v)), \qquad (6)$$

which, by following the same procedure as in Romano [13] and by assuming that the State Equation for the considered flui does not depend on the electric potential, can be shown to be given by

$$R_{min}(U) = (LC_U)^{-1}(p_l^* - p_2). \qquad (7)$$

In conclusion, Theorem 1 holds also for spherically charged drops in which surface tension depends on the mean curvature: the unique modificatio in the statement concerns the definitio of $R_{min}$. Equations (2) lead to

$$R = (LC_U)^{-1}\left(g_l^{-1}(g_v(p_v)) - p_v\right) \qquad (8)$$

which, as $LC_U$ is an increasing function in $U$, allows us to conclude that the equilibrium radius (for a fixe pressure $p_v$) increases with increasing $U$. This is physically understandable when considering the interplay between the repulsive electrostatic forces and the attractive membrane forces due to surface tension.

Finally we remark that: (1) in the presence of a fixe potential $U$ the introduction of Tolman-like corrections leads to smaller equilibrium radii. Indeed, let $L_U(R) := 2\gamma_0/R + U^2/\varepsilon R^2$. Obviously i) $LC_U$ and $L_U$ are both decreasing functions ii) $L_U > LC_U$, and consequently $(L_U)^{-1} > (LC_U)^{-1}$; (2) in the range of the radii for which $\gamma = \gamma_0$ (i.e. $R \gg R_0$) we have :

$$c := g_l^{-1}(g_v(p_v)) - p_v, \qquad \frac{R(U,p_v)}{R(0,p_v)} = \frac{1}{2} + \frac{1}{2}\sqrt{1 + \frac{cU^2}{\gamma_0^2 \varepsilon}},$$

which implies that *the equilibrium (and therefore also the minimal) nucleation radius is larger in the presence of electrostatic fields than without*; this follows simply since

$$\frac{R(U,p_v)}{R(0,p_v)} = \frac{(LC_U)^{-1}(c)}{(LC_0)^{-1}(c)} > 1, \qquad (9)$$

which also holds when $R \simeq R_0$.

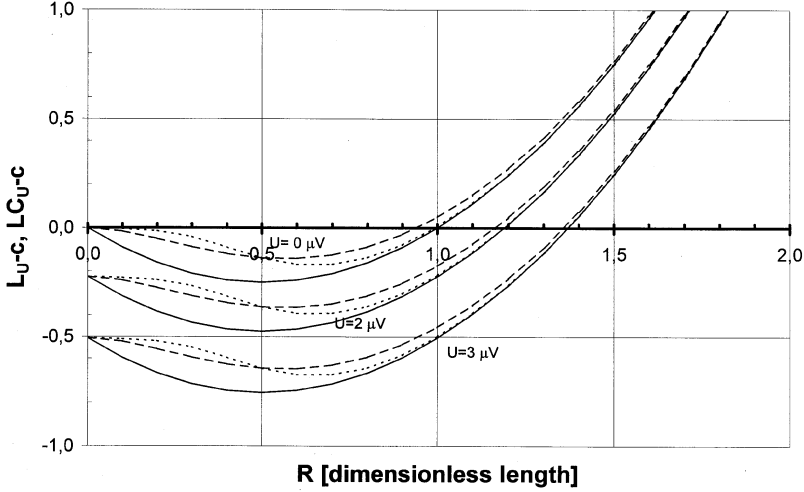

**Fig. 1.** Plot of the functions $LC_U - c$ (see Sect. 2) for three different values of the potential drop ($0\ \mu V, 2\ \mu V, 3\ \mu V$) vs the normalized radial variable. The curves relative to $0\ \mu V$ starts from the origin, the other two start from $-0.25$ and $-0.50$, respectively. The effect of two Tolman-like corrections is shown by dashed curves ($\gamma_1$ large dashes, $\gamma_2$ small dashes). These are compared with the no-correction case ($\gamma_0$ solid curve)

In Table 1 the solution of (8) is shown for $p_v$ in the neighbourhood of the saturation pressure $p_2$ for different values of the electric potential and for the two Tolman-like corrections which we have introduced. For the zero-curvature-surface tension $\gamma_0$ we chose the value for silicon oil, i.e., $\gamma_0 = 0.0201 N/m$, and we set $\alpha = 1$, and $R_0 = 10^{-7} m$.

**Table 1.** Normalized values of $R_{\min}$ as a function of the drop potential for no Tolman-like correction (column 2), correction by (3), i.e., constitutive $\gamma_1$ (column 3) and by (4), i.e., constitutive $\gamma_2$ (column 4).

| $U(\mu V)$ | $R_{\min}(\gamma = \gamma_0)$ | $R_{\min}(\gamma = \gamma_1)$ | $R_{\min}(\gamma = \gamma_2)$ |
|---|---|---|---|
| 0.0 | 1.00000 | 0.94048 | 0.992946 |
| 0.1 | 1.05336 | 1.00707 | 1.048940 |
| 0.3 | 1.36939 | 1.35593 | 1.369050 |
| 0.5 | 1.78650 | 1.78320 | 1.786490 |
| 1.0 | 2.92288 | 2.92278 | 2.922880 |
| 2.0 | 5.26774 | 5.26774 | 5.267740 |

We remark that for potential drops in the range $[0, 2]\ \mu V$ (i) the Tolman-like correction $\gamma_2$ induces variations of the depencence of $R_{\min}$ on $U$ of about 1% and, (ii) the Tolman-like correction $\gamma_1$ induces variations of the depencence of $R_{\min}$ on $U$ of about 7%. In Fig. 1 $L_U - c$ and $LC_U - c$ are plotted as functions of $R$ (normalized to the values $L_0^{-1}(p_l^* - p_2)$) for $c = g_l^{-1}(g_v(p_2)) - p_2$ and several value for $U$. In Fig. 2 we display $R_{min}(U)$ in the three cases: 1) $\gamma = \gamma_0$, 2) $\gamma = \gamma_1$, 3) $\gamma = \gamma_2$. Evidently, for vanishing applied potential drop in the three cases we predict minimum radii that are remarkably different than with such a potential applied. On the other hand, when the applied potential increases, the three cases, as expected, show the same quantitative and qualitative behaviour.

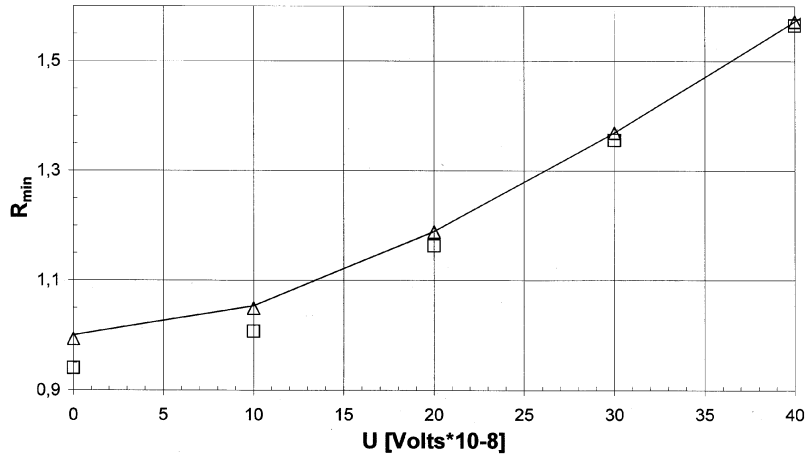

**Fig. 2.** Plots of normalized $R_{\min}(U)$ for different Tolman-like corrections (the data are reported in Table 1): $\gamma = \gamma_0 \rightarrow$ Solid line ; $\gamma = \gamma_1 \rightarrow \square$; $\gamma = \gamma_2 \rightarrow \triangle$

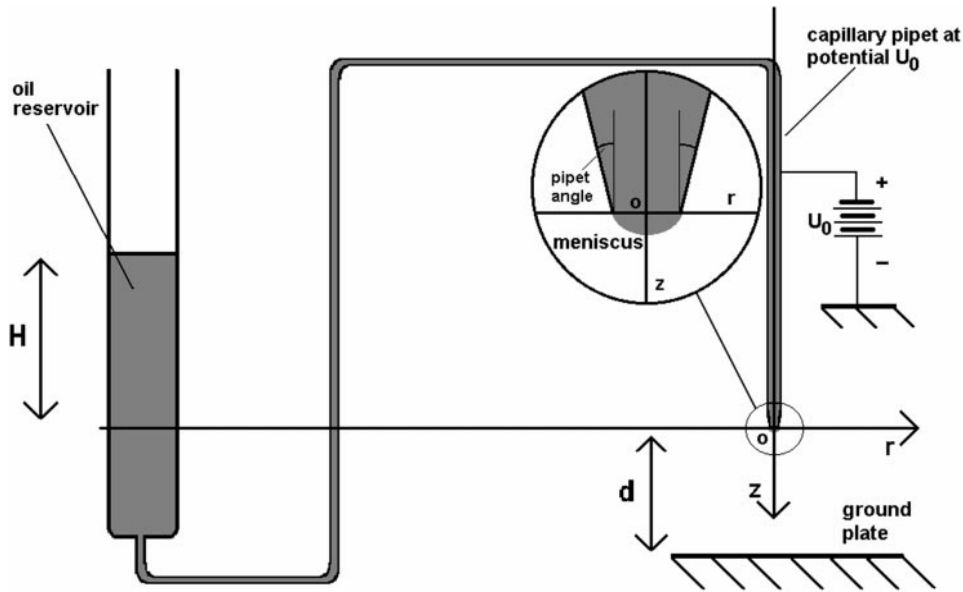

**Fig. 3.** Experimental setup for electrocapillarity experiments. The setup consists of an *oil reservoir* connected to a conducting *capillary pipet*, whose surface is mantained at a piezostatic height H above the end of the capillary pipet where the *meniscus* is formed. The capillary pipet is mantained at a potential $U_0$ whereas a plane capacitive *ground plate* is placed in front of the capillary end. In the inset is shown the geometry of the capillary end: the meniscus is pendant from a truncated conical pipet with a given *pipet angle*. It is assumed that the pipet angle is chosen to match the meniscus of the pipet in a relatively smooth way (so no relevant concentration of surface charge appears at the separation line between pipet and meniscus)

## 3 Electrocapillary menisci

### 3.1 Basic equations

We conceive a system as schematically sketched in Fig. 3. A conducting truncated conical pipet is placed in front of a plane capacitive plate. A conducting liquid is pushed by a piezostatic pressure to form a capillary meniscus suspended at the lower opening cross section of the pipet.

The range of the considered potential drop between the pipet and the plate will be chosen as a function of the angle of the pipet: it will be required that at the edge between the surfaces delineating the pending meniscus and the pipet the discontinuity of the normals induces a concentration of surface charge which is significantl smaller than the surface charge concentrated at the tip of the meniscus.

The motivation of this choice lies in the simplificatio which it implies in the study of the electric fiel generated between the pipet and the plate. In particular, no significan electric-fiel concentration appears in the neighbourhood of the aforementioned edge. As a practical consequence, if the theory should be tested experimentally, a family of capillary tubes with different cone angles but with fixed inferior basis would have to be considered in order to allow measurements involving a large range of applied potential drops.

In the region $D$ between the plate and the meniscus the electric potential satisfies Laplace's equation. We explicitly remark here that in solving the elliptic problem which follows we consider the region $D$ as constituted by the part of the space between two infinite surfaces representing respectively the conducting surface of the pipet and the grounded plate. The electric field in the pending liquid vanishes. The surface force balance, generalizing the capillary Laplace law for surface tension, determines the shape of the pending meniscus $S$. Scaling all the lengths by the capillary radius, $R$, the potential by $U_0$, i.e., the chosen drop of the potential between the plate and the pipet, the surface tension by $\gamma_0$, i.e., the surface tension for plane interfaces, and the charge density, $\sigma$, by $\sqrt{R/(\varepsilon \gamma_0)}$ (this choice naturally arises when recalling the considerations about the function $LC_U$ introduced in the previous section) the aforementioned basic equations, under the assumption of cylindrical symmetry for the considered apparatus, read

$$\Delta U = 0, \quad \text{in } D, \tag{10}$$

$$k_1(k_2 + z(r)) + \frac{\gamma(R_g)}{R_g} + \frac{\sigma^2}{2} = 0, \quad \text{on } S, \tag{11}$$

where

$$k_1 := \rho g R^2 / \gamma_0, \quad k_2 := H/R, \quad \sigma = \frac{E}{\varepsilon}; \tag{12}$$

$\rho$ is the liquid mass density, $H$ the piezostatic height, $\Delta$ the Laplace operator, $z(r)$ the function specifying the profile of the meniscus in the cylindrical coordinate system of which the axes are oriented as shown in Fig. 3: the $z$-axis is pointing in the direction of gravity from the pipet to the grounded plate, and $E$ is the component of the electric field orthogonal to the meniscus.

In the present section we analyze the effect of Tolman's corrections (3) and (4) as well as those due to:

$$\gamma_3(R_g) = \gamma_0 \frac{1}{2} \left[ 3 - \tanh\left(\frac{R_g - R_0}{R_0}\right) \right]. \tag{13}$$

While (3) and (13) show an increasing behaviour as functions of $R_g$, (13) is decreasing. These three Tolman type corrections are able to efficiently capture their most important features. The boundary conditions are

1) for the potential

$$U = 1 \tag{14}$$

on the meniscus and the pipet, and

$$U = 0 \tag{15}$$

on the plate which is at a distance $d$ from the pipet, and, because of the cylindrical symmetry,

$$\frac{\partial U}{\partial r}(r = 0, z) = 0, \quad z \in [0, d] \tag{16}$$

2) for the $z(r)$ function we have ($r = 1$ represents the edge of the lower cross section of the pipet)

$$z(1) = 0, \quad z'(0) = 0. \tag{17}$$

The first condition represents the patching condition between the meniscus and the lower cross section of the pipet, while the second is a consequence of the cylindrical symmetry of the pipet.

This is a system of partial differential equations (PDE) with "free" unknown boundary (the surface $z(r)$) on which a Dirichlet condition on the potential is assigned. In fact, by varying $U_0$, the free boundary will change the shape because it is determined by the electric field in the region $D$. The problem can be classified as a Dirichlet problem for the potential coupled with a Sturm-Liouville (Two Boundary Values) problem for the profile function $z(r)$.

In order to clearly specify the range of applicability of (10), (11) the following remarks are in order:

1) We suppose $U_0$ to be sufficientl small that, at the meniscus tip (i.e. at the location of maximum surface charge density), no charge emission takes place; so the Laplace equation for the potential can be employed instead of the Poisson equation.
2) It is assumed that the only effect due to the internal pressure of the liquid, i.e., the cohesive molecular attraction due to the $a/V^2$ terms in the equation of state, is the presence of the surface tension; this obviously implies the well-known anomaly of the interface stability, as $g \to 0$.
3) We do not consider cylindrical capillary tubes for the reason that in this case the potential has two singular points on the attachment between the meniscus and the capillary tube (see Joffre [14]). We prefer to deal with conically shaped capillary tubes, that is with a system where the maximum radius of curvature and the maximum electrostatic surface-charge density are on the "top" of the meniscus. In this way we avoid the presence of corners, which introduce strong singular points for the electric fiel and complicate the problem of singling out the effects of Tolman-like corrections. Indeed, in the apparatus which we introduce, the surface formed by the union of the meniscus and the outer surface of the pipet is a smooth equipotential surface.

*3.2 Variational method for menisci of parabolic shape*

In the literature, it is usually assumed that the menisci have a rotationally symmetric parabolic shape. In this paper, we fin the shape of the meniscus by solving, via a mixed numerical-analytical method, the system (13), (14). In addition, in order to validate this mixed method, the system (13), (14) for $\gamma(R) = \gamma_0$ is also solved by a variational method by assuming that the meniscus has a parabolic shape parametrized by its height $z_m$. This can be done by minimizing the free energy as a function of $z_m$ when a parabola is chosen as a test-function.

The free energy expression is composed of three parts: the gravitational energy, $\epsilon_g$, the energy due to the surface tension, $\epsilon_\gamma$, the electrostatic energy plus the work done by a generator keeping the tube-meniscus system to a fixe drop of potential, $\epsilon_e$, viz.

$$\epsilon = \epsilon_g + \epsilon_\gamma + \epsilon_e, \tag{18}$$

where

$$\epsilon_g = -\int_V \rho_l g dV, \quad \epsilon_\gamma = \int_S \gamma ds, \quad \epsilon_e = \frac{\varepsilon}{2} \int_D E^2 dV. \tag{19}$$

$V$ and $S$ are the volume and the surface of the meniscus respectively, and $D$ is the region in which the electric fiel is not negligible. The family of test functions that minimizes (18) is constructed in Appendix 1 as is the critical value for the height $z_m$. In so doing the expressions for the gravitational and surface energies are identical to those given in Joffre [14]. Equation (19), on the other hand includes both aforementioned electric terms.

For the electrostatic energy we use the expression

$$\epsilon_e = 2\pi\varepsilon U_0^2 \left(\frac{R_{eff}}{R}\right)^2 \frac{z_m}{\ln[1 + \frac{4z_m(d-z_m)}{R^2}]}, \tag{20}$$

which is derived in Appendix 1 and motivated in Sect. 4 below. Here, $R$ is the radius of the lower basis of the truncated conical pipet, $d$ is the distance between this last basis and the plate and $R_{eff}$ is define in Appendix 1. There, the electric fiel between the plate and the conducting pipet is estimated by assuming both to extend to infinity This assumption does not introduce relevant errors in the local estimate of the electric fiel in the neighbourhood of the meniscus needed in (11). However, in the global variational energy estimate (19) one must account for the finit amount of energy of the electric fiel generated by the potential drop, $U_0$, between the plate and the pipet the extensions of which are finite This is exactly done by introducing $R_{eff}$ which is determined via a least square adjustment to numerical curves and is seen to have a value of about $R/3$.

## 4 An approximate solution of (10) and (11)

Our procedure to calculate the menisci profil is based on the steps described in the following subsections.

*4.1 Numerical procedure to integrate (11)*

After having found for (11) an equivalent expression in normal form (see Sect. 5.2 for a more detailed discussion on the subject) we integrate (14) by the standard Newton integrator supplied in *Mathematica*©. A shooting procedure is enployed based on the Cauchy initial values given by the estimate $z(0) = z_m$ and the condition $z'(0) = 0$ imposed by the cylindrical symmetry. The $z_m$ values are then adjusted by a sequence of integrations until the condition $z(R) = 0$ is verifie within an error less than $10^{-4}$. The accuracy of the method is limited by the machine precision, which in single precision is $10^{-8}$. We fin that this precision is sufficien to catch the main geometrical features of the menisci.

The robustness of the shooting method was further tested by the following two alternatives: (i) We used as Cauchy initial data $z(R) = 0$, $z'(R) = z'_R$ and tuned this last value until the boundary condition $z'(0) = 0$ was verified The solution and the boundary conditions in the two alternative shootings are reproduced within the accuracy of the integration routine. (ii) Both shooting procedures were tested by starting from the corresponding arrival boundary point and values and integrating back to the initial point. The success of this test is a necessary condition to the stability of the employed numerical procedure.

*4.2 Approximate analytical determination of the electric field on the surface of the meniscus*

In (11), there appears the value of the electric fiel on the surface of the meniscus. In the numerical procedure described in the previous section an approximate analytical expression is used that is obtained as follows: we solve in the firs step (10) in parabolic coordinates. As the shape of the meniscus is very similar to a parabola and the plate can be regarded as a distant and fla parabola we use firs step to estimate the electric potential in the neighbourhood of the meniscus. In other words we surmise that the potential generated by the meniscus has the same *functional form*, as generated by a locally-close parabolic meniscus. Indeed, even if the meniscus does not exactly possess parabolic shape the potential on the $z(r)$ surface is still given as a function of $z$ by (28) below, and this expression will be used, in a second step, when solving (11).

This firs step can be considered the initial block of an iterative method in which the new shape of the meniscus can be used to calculate a new potential and so on; the convergence of this method is so fast that it determines the limit potential after only one iteration. This we have checked numerically via the estimation of the second iteration correction which is two orders of magnitude smaller.

4.2.1 First step: Solution of (10) in parabolic coordinates

The electric fiel generated by two homofocal parabolas (Durand [17]) is

$$E(\zeta, \eta) = \frac{k}{\sqrt{\zeta^2 + \eta^2}} \cdot \frac{1}{\zeta}, \tag{21}$$

where $\eta$ and $\zeta$ are the parabolic coordinates which are related to the cylindrical coordinates by

$$z = \frac{\zeta^2}{2} - \frac{\eta^2}{2}, \quad r = \eta\zeta. \tag{22}$$

Equation $\zeta = \zeta_1$ describes the parabola approximating the outer surface of the pipet and meniscus. Its vertex $V_1$ has cylindrical coordinates $(0, -z_m) := (0, -\zeta_1^2/2)$.

Approximating the meniscus as a parabola with equation $\zeta = \zeta_1$ and the plate as a parabola with equation $\zeta = \zeta_0$ with vertex $V_0$ the cylindrical coordinates of which are $(0, -d) := (0, -\zeta_0^2/2)$, one can easily deduce from (29) (see Durand [17])

$$k = -\frac{U_0}{\ln(\zeta_1/\zeta_0)}. \tag{23}$$

4.2.2 Second step : Approximation for the electric field close to the meniscus

The assumption that the form of the plate is seen by the meniscus as a parabola is equivalent to the inequality

$$\sqrt{2d} \gg \sqrt{2z_m} \tag{24}$$

so that when estimating the electric field close to the meniscus the parabola with vertex $V_0 \equiv (0, -\zeta_0^2/2) = (0, -d)$ can with sufficient approximation be replaced by a plane.
The electric field on the first parabola is given by

$$E(\zeta = \sqrt{2z_m}, \eta) = -\frac{\frac{U_0}{\ln(z_m/d)^{1/2}}}{\sqrt{2z_m + \eta^2}} \cdot \frac{1}{\sqrt{2z_m}}, \tag{25}$$

or in view of (29), (30), since

$$\eta^2 = \sqrt{z^2 + r^2} - z, \tag{26}$$

$$E(z,r) = -\frac{\frac{U_0}{\ln(z_m/d)^{1/2}}}{\sqrt{2z_m + \sqrt{z^2 + r^2} - z}} \cdot \frac{1}{\sqrt{2z_m}}. \tag{27}$$

This last equation can be considered as a local approximation for the electric field close to the meniscus also when its shape, remaining close to the parabolic one, is not exactly parabolic. If the shape of the meniscus in cylindrical coordinates is given by the equation $z = z(r)$ we are allowed to still use the obtained functional form for the field i.e.

$$E(z,r) = -\frac{\frac{U_0}{\ln(z_m/d)^{1/2}}}{\sqrt{2z_m + \sqrt{z(r)^2 + r^2} - z(r)}} \cdot \frac{1}{\sqrt{2z_m}}. \tag{28}$$

This is the expression of the electric field in which the form $z(r)$ of the meniscus and its height $z_m$ at the center appear explicity. We will use it to solve (11), which thus becomes an ordinary differential equation for the unknown function $z(r)$.

## 5 Discussion of the numerical results

### 5.1 The case of "large radius" capillary tube

For a capillary tube with radius much greater than $R_0$ we can neglect the Tolman-like correction and solve the system of equations in which $\gamma(R_g) = \gamma_0$. In Fig. 4a the results of the integration for different values of the potential are shown; in the same figure the parabolic shapes corresponding to the same height of the meniscus are also shown. The values of the parameters are listed in Table 2.

**Table 2.** Values of parameters used for quantitative evaluations

| |
|---|
| $H = 2.00 \cdot 10^{-3}$ m |
| $\gamma_0 = 2.01 \cdot 10^{-2}$ N/m |
| $d = 10^{-2}$ m |
| $\rho_l = 930.00$ $Kg/m^3$ |
| $R_0 = 1.49 \cdot 10^{-7}$ m |

We observe that for high values of the potential the menisci become more "fat" i.e., more concave with respect to the parabolically shaped menisci. Note that the maximum potential (limit potential), above which the meniscus is unstable, is reached when $z_m = R$. This is a well-known property of pending menisci: Our choice of the geometry of the pipet allow us to find all shapes up to the theoretical limit value. In Fig. 4b the surface charge density for all calculated shapes of the meniscus is shown. The maximum of the surface charge density occurs at the tip of the meniscus as expected. The numerical values are very close to those found by Joffre [14] in a similar case. In Table 3a the calculated values for $z_m$ are listed for varying $U_0$. In Fig. 4c the meniscus height $z_m$ is plotted vs. the applied drop of the potential as estimated with both the variational method described in Sect. 3.2 and the two-step method described in the previous section. The best-fi test shows that the plotted dependence is very close to be quadratic when the more accurate second method is used.

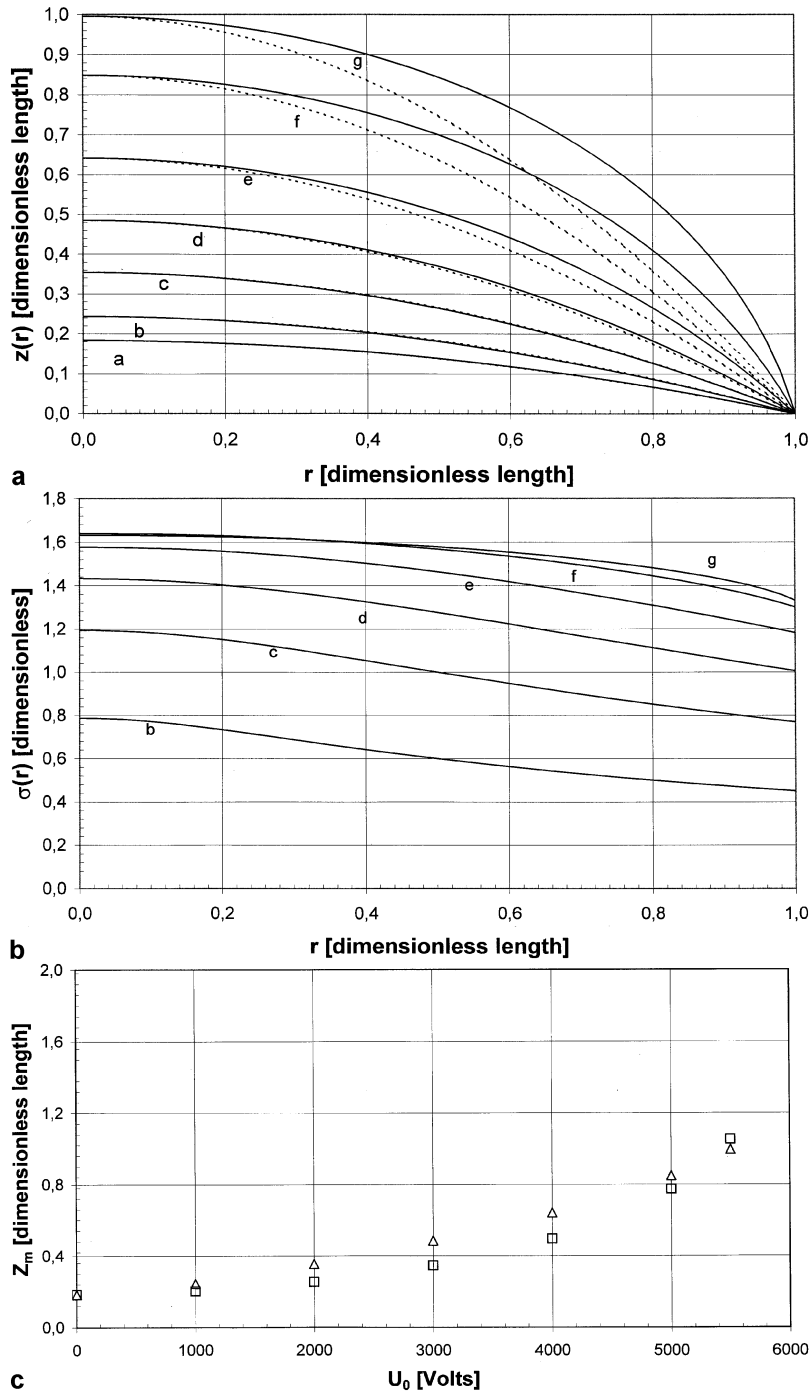

**Fig. 4.** Large radius case; the parameters of the simulations are listed in Table 2: **a** Profile of the meniscus for the values of a potential drop (a-g) as listed in Table 3a, using (29) (solid line). In the same figur is also shown the parabolic approximation (dashed line starting at the same values of normalized height $z_m$) for the same values of potential drop, **b** normalized surface charge density for the same potentials (a-g) as in **a**, **c** normalized meniscus height vs applied potential found using the variational method (□) and using (29) (△)

In concluding the comparison of our results with those found by Joffre [14] we explicitly note, however, that his limit potential, and consequently his limit height, was found to be smaller than the theoretical limit owing to the different solution adopted for the tube-menisci system. Most likely, our solution is smoother for the truncated conically shaped pipet, because it is not affected by an edge singularity in the surface charge.

*5.2 The case of "small radius" capillary tube*

In Sect. 5.1 we were allowed to use $\gamma(R_g) = \gamma_0$. Instead, when the mean curvature radius $R_g$ is comparable with the radius $R_0$ Tolman-like corrections should be introduced: this is the case of "small" radius capillaries.

We consider all material parameters having the same value as in the previous section but the capillary radius is now $10^{-3}$ times smaller. We consider therefore (3) and (4) with the value of $R_0 = \frac{1}{5}R$ and $\alpha = 1$ (see dell'Isola et al. [11]). These choices are surely physically reasonable at least in the neighbourhood of the critical temperature and/or for fluid with large molecules (e.g. for polymeric fluids)

**Table 3a.** Normalized values of $z_m$ as function of drop potential for a "large capillary radius" ($R = 0.745 \cdot 10^{-3} m$)

|   | $U_0(V)$ | $z_m$ |
|---|---|---|
| a | 0 | 0.1834 |
| b | 1000 | 0.2436 |
| c | 2000 | 0.3542 |
| d | 3000 | 0.4848 |
| e | 4000 | 0.6412 |
| f | 5000 | 0.8488 |
| g | 5500 | 0.9958 |

Tolman-like corrections transform (11) into an ordinary differential equation which is not in normal form. Indeed, (11) takes the form

$$F(z, z', z'') = H(z), \quad ()' = \frac{d}{dr}, \tag{29}$$

where $H$ is the function of $z$ obtained by summing up the firs and third addend in (11) and $F$ is a trascendental function of the argument $z''$. This circumstance represents a remarkable difficulty to overcome it one can try two methods. In the first we could try to numerically solve by a finit difference scheme the trascendent algebraic equation $F(z_0, z'_0, z'') = H(z_0)$ for every choice of initial data $(z_0, z'_0)$. This method proved to be poorly convergent near the top of the menisci, where we expect to fin the most interesting physical features of the studied system. In the second, we solved (29) iteratively by findin at step $n$ the solution of the problem

$$F(z_{n-1}, z'_{n-1}, z''_n) = H(z_n). \tag{30}$$

This methods demonstrates a very efficien convergence when $z_0$ is chosen to be of parabolic shape. Indeed with this choice $z_1(r)$ and $z_2(r)$ are nearly coinciding within an error of approximately 2%. Therefore, the Tolman-like corrections can be transformed from a functional operator depending on $z(r)$ into a known function $T_i(r, U_0) := \gamma_i(R_g(r, z_0(r))$ $i = 1, 2, 3$. Here, $R_g(r, z_0(r))$ denotes the mean radius of curvature corresponding to the meniscus the shape of which is given by $z_0(r)$ of the variable $r$ by simply replacing $z(r)$ by the best parabolic fi corresponding to the potential drop $U_0$.

In Fig. 5 the functions $T_i(r, U_0)$ are shown as functions of $r$ for different values of $U_0$; it must be underlined that, as expected, the differences $1 - T_i(r, U_0)/\gamma_0$ attain their maximum at the top of the menisci. Using the parameters listed in Table 2 we display in Fig. 6 for different values of $U_0$ the shapes of the menisci with and without the Tolman-like corrections. We note that the firs two Tolman-like corrections produce slightly different profile for the menisci: for a fixe potential $U_0$ the height of the meniscus is bigger for the correction (4). For the third correction the effect is qualitatively opposite. We do not show the surface charge density plots because in the considered case they are qualitatively very similar to those of the large radius case.

In Table 3b we list the calculated values for $z_m$ for variations of the value of $U_0$ and different Tolman-like corrections.

In Fig. 7 we plot the height $z_m$ of the equilibrium meniscus as a function of the potential drop $U_0$ for constant $\gamma$ and for the three Tolman-like corrections. We remark that: (1) for the chosen material parameters the firs two Tolman-like corrections show the same qualitative and nearly the same quantitative behaviour; (2) the difference between the estimated $z_m$ in the case $\gamma = \gamma_0$ and in the case of Tolman-like corrections is an increasing function of $U_0$; (3) for the firs two Tolman-like corrections the limit equilibrium drop of potential (i.e. the potential corresponding to $z_m = 1$, in normalized units, case in which the tip height equals the pipet radius and the theoretical maximal height of the meniscus tip is attained) is smaller in the case of Tolman-like corrections than in the case $\gamma = \gamma_0$; the absolute value (with our choice of material parameters) of this difference amounts to 40 $V$, corresponding to a decrease of an amount of 5%; this effect can be easily shown by electric measurements and is physically evident: indeed owing to the "weakening" of surface tension, especially at the top of the menisci, because of Tolman-like corrections a greater height with respects of the case $\gamma = \gamma_0$ is reached; consequently the equilibrium limit potential is found lower. (4) For a length scale effect the limit drop of potential decreases as a function of the capillary radius (for more details about this point see dell'Erba [18]); (5) as in the case of large capillary radii the dependence of $z_m$ on $U_0$ is very close to a quadratic one.

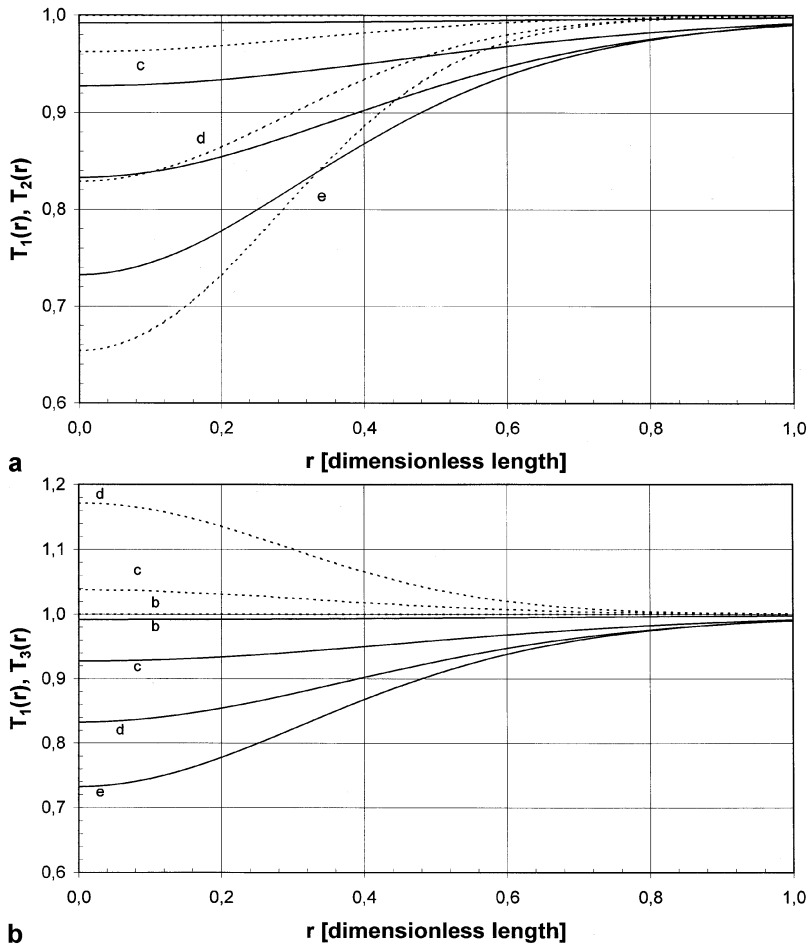

**Fig. 5.** Tolman corrections, the parameters of the simulations are listed in Table 2: **a** $T_1$ (solid line) and $T_2$ (dotted line) for small radius vs normalized radial variable, only the relevant corrections of the set (a-e) listed in Table 3b are shown, **b** $T_1$ (solid line) and $T_3$ (dotted line) for small radius vs normalized radial, only the relevant corrections of the set (a-e) listed in Table 3b are shown

**Table 3b.** Normalized values of $z_m$ as a function of the potential drop for a "small capillary radius", in the case of no Tolman-like correction (column 2), correction by (3) (column 3) by (4) (column 4) and finall by (13) (column 5); ($R = 0.745 \cdot 10^{-6} m$)

|   | $U_0(V)$ | $z_m^0$ | $z_m^1$ | $z_m^2$ | $z_m^3$ |
|---|---|---|---|---|---|
| a | 0 | 0.00017 | 0.00017 | 0.00017 | 0.00017 |
| b | 200 | 0.26020 | 0.26100 | 0.26040 | 0.26020 |
| c | 400 | 0.47700 | 0.48720 | 0.48080 | 0.46420 |
| d | 600 | 0.69910 | 0.72860 | 0.71960 | 0.61330 |
| e | 800 | 0.94840 | 1.00000 | 0.99520 | 0.67520 |
|   | 840 | 1.00000 | - | - | 0.68520 |

## 6 Conclusions

Tolman [1], using the Gibbs' excess theory, deduced a formula relating the surface tension to the equilibrium radii for droplets and bubbles that are immersed in their own vapour. La Mer and Pound [3] proved that available experimental evidence was in contradiction with Tolman's theoretical prediction. Fisher and

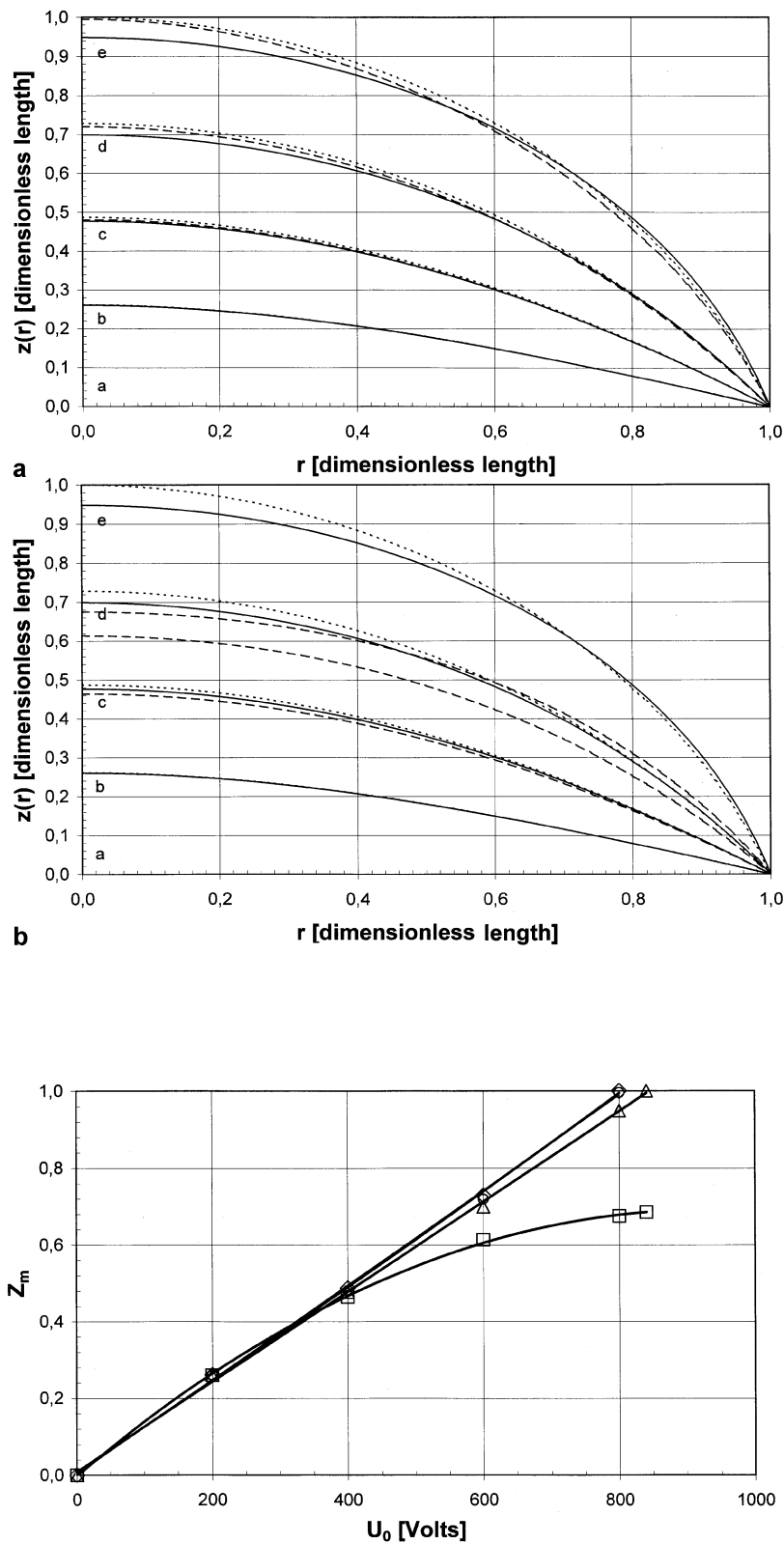

**Fig. 6.** Small radius case: **a** Profile of the meniscus for the values of potential drop (a-e) listed in Table 3b for the small radius case: with Tolman-like correction $\gamma = \gamma_1$ (dotted line), for Tolman-like correction $\gamma = \gamma_2$ (dashed line), and without correction $\gamma = \gamma_0$ (solid line), **b** Profile of the meniscus for the values of potential drop (a-e) listed in Table 3b for the small radius case: with Tolman-like correction $\gamma = \gamma_1$ (dotted line), for Tolman-like correction $\gamma = \gamma_3$ (dashed line), and without correction $\gamma = \gamma_0$ (solid line)

**Fig. 7.** Normalized meniscus height $z_m$ vs applied potential found using $\gamma = \gamma_0$ (△), $\gamma = \gamma_1$ (◇), $\gamma = \gamma_2$ (○) and $\gamma = \gamma_3$ (□)

Israelachvili [4] designed a more precise experimental apparatus which showed for microscopic curved interfaces a dependence of the surface tension on curvature. Alts and Hutter [6], using the modern continuum mechanics approach firs made plausible the constitutive dependence of surface tension on interface curvature. Dell'Isola et al. [11] using second gradient 3-D theories, proved that equilibrium surface tension must depend on curvature simply as a result of the second law of thermodynamics and proposed, starting from the equations of state, some equilibrium relations for macroscopic surface tension as a function of curvature (see (3) and (4)).

In this paper we prove that a Tolman-like dependence of the surface tension on mean curvature has effects on the shape of some equilibrium electrically charged vapour-flui interfaces which could be measurable. Indeed for electrically conducting materials we propose to use an experimental apparatus in which the limit equilibrium drop of the electric potential (before instability) depends notably on the Tolman-like correction for surface tension and, in particular, on its dependence on the mean curvature. This experimental apparatus seems suitable to detect efficientl the searched physical effect, as it is based on the measurements of the electrical potential drop; these measurements can much more precisely be conducted than purely mechanical ones, due to the refine electronic equipment now available. Moreover, we fin how the equilibrium radius of charged drops depends on their surface charge.

Possible developments of our results could concern (1) a more refine numerical analysis for the prediction of the behaviour of the presently proposed experiment; (2) the study of menisci with charge emissions; (3) the study of instabilities and rupture profiles (4) the study of the effect on surface tension and shape of menisci of the presence of impurities in one or both present phases.

*Acknowledgement.* We thank Prof. A. Di Carlo and Prof. K. Hutter for fruitful discussions and their friendly criticism.

**Appendix 1**

In Sect. 3.2 we introduced a variational formulation of the problem of the determination of equilibrium menisci. This formulation is essentially based on the representation formula (20) for the electrostatic energy: in the present Appendix we justify it.

We consider parabolic and cylindrical coordinates $(\zeta, \eta)$ and $(r, z)$ respectively. The coordinate lines of the parabolic system are parabolas with axis of simmetry $r = 0$. The coordinate transformation is given by

$$z = \frac{\zeta^2}{2} - \frac{\eta^2}{2}, \quad r = \eta\zeta. \tag{31}$$

We now consider the family $\mathcal{F}$ of parabolas given by the equation

$$z(r) = z_m \left(1 - \frac{r^2}{R^2}\right) + h(z_m) \tag{32}$$

parametrized by the variable $z_m$. We shall fin the expression for $h(z_m)$ by assuming that every element of $\mathcal{F}$ is a coordinate curve of the introduced parabolic system of coordinates. Therefore, $\mathcal{F}$ can be parametrized also in terms of $\zeta_0$ as follows

$$\zeta = \zeta_0 \quad \text{and} \quad \eta \text{ variable}. \tag{33}$$

We now request that the parametrization (32) is equivalent to (33) and obtain the set of equalities

$$\eta = 0 \Rightarrow \frac{\zeta_0^2}{2} = z(0) = z_m + h(z_m), \tag{34}$$

$$r = \zeta_0 \eta \Rightarrow \eta^2 = \frac{r^2}{\zeta_0^2} = \frac{r^2}{2(z_m + h(z_m))}, \tag{35}$$

$$z(r) = z_m + h(z_m) - \frac{\eta^2}{2} = z_m + h(z_m) - \frac{r^2}{4(z_m + h(z_m))} = z_m + h(z_m) - \frac{z_m r^2}{R^2}, \tag{36}$$

where the last equality must hold for every $r$. As a consequence we obtain

$$h(z_m) = \frac{R^2}{4z_m} - z_m. \tag{37}$$

Therefore, using parabolic coordinates, we have

$$\zeta_0^2 = 2(z_m + h(z_m)) = \frac{R^2}{2z_m}, \qquad \eta^2 = \frac{2z_m r^2}{R^2}. \tag{38}$$

We now assume that, for every $z_m$, the parabola representing the meniscus is placed in front of a parabola, representing the plate, which is a distant $d$ apart from it in the increasing direction of the $z$-coordinate. Using the solution in parabolic coordinates of the Laplace electrostatic problem found in Durand [17]

$$E(\zeta_0, \eta) = \frac{k}{\sqrt{\zeta_0^2 + \eta^2}} \cdot \frac{1}{\zeta_0}, \tag{39}$$

we find that the normal component of the electric field to the meniscus is

$$E(\zeta_0 = \frac{R^2}{2z_m}, \eta) = \frac{k}{\sqrt{\frac{R^2}{2z_m} + \eta^2}} \cdot \frac{1}{\frac{R^2}{2z_m}}, \tag{40}$$

where, as the plate is placed at a distance $d$ from the meniscus,

$$k := \frac{2U_0}{\ln\left((d + h(z_m))/(z_m + h(z_m))\right)}. \tag{41}$$

We are now ready to estimate the electric field energy $\epsilon_e$: we start by remarking that in our case it is given by

$$\epsilon_e = \frac{QU_0}{2} \tag{42}$$

where $Q$ is the total surface charge on the union of the pipet and the meniscus and is given by

$$Q = \int_{S \cup P} \varepsilon E \, ds, \tag{43}$$

where $E$ is the component of the electric field normal to the surface and $P$ is the pipet surface.

To estimate this last integral we remark that the field given by (40), is a good approximation for the normal component of the electric field close to the meniscus. However as the field (40) is generated by an infinite distribution of charge while the considered system is constituted by finite conducting surfaces it represents an overestimation of the field $E$. Therefore, neglecting the charges on the pipet, we can introduce an effective radius $R_{eff}$ by means of the equality

$$Q =: \int_0^{R_{eff}} \int_0^{2\pi} \varepsilon E J(r, \vartheta) dr d\vartheta \tag{44}$$

in which $J(r, \vartheta)$ is defined by $ds =: J(r, \vartheta) dr d\vartheta$ and $E$ is given in (40). As a consequence

$$Q = 2\pi\varepsilon \, k z_m \left(\frac{R_{eff}}{R}\right)^2, \tag{45}$$

in which $R_{eff} < R$ because edge effects lower the total charge on the considered finite parabolic conductor relative to the total charge present on the same finite conductor regarded as a part of an infinite one. By replacing (45) in (42) we obtain

$$\epsilon_e = 2\pi\varepsilon U_0^2 \left(\frac{R_{eff}}{R}\right)^2 \frac{z_m}{\ln[1 + \frac{4z_m(d-z_m)}{R^2}]} . \tag{46}$$

A simple test of the validity of (46) is obtained if it is applied to plane circular menisci with very large radii. This is done by estimating the limit

$$\lim_{z_m \to 0} \epsilon_e = \frac{1}{2} C U_0^2, \quad \text{with} \quad C = \frac{\pi R^2 \varepsilon}{d}, \tag{47}$$

which gives a well-known classical result.

For every $R_{eff}$ we thus obtain a dependence of $z_m$ on $U_0$. Among the curves in this family we choose that one which better fit the curve obtained with the method developed in Sect. 4. In this way we obtain

$$R_{eff} \approx 0.31 \, R. \tag{48}$$

## References

1. Tolman RC(1949) The effect of droplet size on surface tension. J.Chem.Phys. 17, n.3, p.333
2. Gibbs JW (1948) The Scientifi Papers of J.W.Gibbs. Dover, New York
3. LaMer VK, Pound GM (1949) Surface Tension of small Droplet from Volmer and Flood's nucleation data. Chem.Phys.Lett. 17, p.1337
4. Fisher LR, Israelachvili JN (1980) Determination of the capillary pressure in menisci of Molecular Dimension. Chem.Phys.Lett. 76, n.2, p. 325
5. Defay R, Prigogine I (1951) Tension superficiell and adsorption. Eds.Desoer, Liege
6. Alts T, Hutter K (1988-89) Continuum description of the dynamics and thermodynamics of phase boundaries between ice and water, Part I,II,III,IV. J. NonEq. Therm. 13, 14
7. dell'Isola F (1994) On the lack of structure of Defay-Prigogine 2D-continua. Arch.Mech. 46, 3, 329
8. Germain P (1972) La méthode des puissances virtuelles en mécanique des milieux continus. Premiere partie: Théorie du second gradient. J.de Mecanique 12, n.2, p. 235
9. dell'Isola F, Rotoli G (1995) Validity of Laplace formula and dependence of surface tension on curvature in second gradient fluids Mech.Res.Comm. 22,n.5,p.485-490
10. dell'Isola F, Gouin H, Seppecher P (1995) Radius and Surface Tension of Microscopic bubbles by second gradient theory. C.R.Acad.Sci. Paris, 320, IIb, p.211-217
11. dell'Isola F, Gouin H, Rotoli G (1996) Nucleation of spherical shell-like interfaces by second gradient theory: numerical simulations. Eur.J.Mech.B/Fluids 15, 4, 545
12. dell'Isola F, Kosinski W (1993) Deduction of thermodynamic balance laws for bidimensional nonmaterial directed continua modelling interphase layers. Arch.Mech. 45, 3, 333
13. Romano A (1983) Properties of the Gibbs Potential and the Equilibrium of a liquid with its Vapour. Rend.Sem.Mat.Univ.Padova 69, 259
14. Joffre GH (1984) Profil de Menisques Electrises avec ou sans Emission de Charges. Thèse de doctorat d'etat, Universite' Pierre e Marie Curie (Paris VI)
15. Kumar FJ, Jayaraman D, Subramanian C, Ramasamy P (1991) Curvature dependence of surface free energy and nucleation kinetics of $CCl_4$ and $C_2H_2Cl_4$ vapours. J. of Material Sci. Lett. 10, p.608-610
16. dell'Isola F, Romano A (1987) A phenomenological approach to phase transition in classical fiel theory. Int.J. En.gng.Sci. 25, 11/12 pp.1469-1475
17. Durand E (1966) Electrostatique, Tome II. Masson et C. Editeurs
18. dell'Erba R (1997) Elettrodinamica Razionale dei Continui. Tesi di dottorato in Meccanica Teorica ed Applicata Università di Roma LA SAPIENZA